\documentclass[twocolumn, superscriptaddress,nofootinbib,nofootinbib]{revtex4-1}   
\usepackage{graphicx}  
\usepackage{dcolumn}   
\usepackage{bm}        
\usepackage{amssymb, amsmath}
\usepackage[utf8]{inputenc}

\def\beq{\begin{equation}}
\def\eeq{\end{equation}}
\usepackage[usenames,dvipsnames,svgnames]{xcolor}
\usepackage{hyperref}
\hypersetup{colorlinks, citecolor=black, linkcolor=black, urlcolor=black}

\usepackage{amsmath}
\usepackage{amsfonts}
\usepackage{amssymb}
\usepackage{bbm}                
\usepackage{graphicx, rotating}
\usepackage{epstopdf}
\usepackage{epsfig}
\usepackage{latexsym}
\usepackage{graphicx}
\usepackage{bbold}
\usepackage{color}
\usepackage{amsmath,amssymb}

\newcommand{\units}[1]{\textrm{\,#1}}

\begin{document}
\title{Signs of Dark Matter at 21-cm?}
\author{Rennan Barkana}
\affiliation{Raymond and Beverly Sackler School of Physics and Astronomy, Tel-Aviv University, Tel-Aviv 69978, Israel}
\author{Nadav Joseph Outmezguine}
\affiliation{Raymond and Beverly Sackler School of Physics and Astronomy, Tel-Aviv University, Tel-Aviv 69978, Israel}
\affiliation{School of Natural Sciences, Institute for Advanced Study,
Einstein Drive, Princeton, NJ 08540, USA}
\author{Diego Redigolo}
\affiliation{Raymond and Beverly Sackler School of Physics and Astronomy, Tel-Aviv University, Tel-Aviv 69978, Israel}
\affiliation{School of Natural Sciences, Institute for Advanced Study,
Einstein Drive, Princeton, NJ 08540, USA}
\affiliation{Department of Particle Physics and Astrophysics, Weizmann Institute of Science, Rehovot 7610001,Israel}
\author{Tomer Volansky}
\affiliation{Raymond and Beverly Sackler School of Physics and Astronomy, Tel-Aviv University, Tel-Aviv 69978, Israel}
\affiliation{School of Natural Sciences, Institute for Advanced Study,
Einstein Drive, Princeton, NJ 08540, USA}

\begin{abstract}
Recently the EDGES collaboration reported an anomalous absorption signal in the sky-averaged  21-cm spectrum around $z=17$.   
Such a signal may be understood as an indication for an unexpected cooling of the hydrogen gas during or prior to the so called Cosmic Dawn era.  Here we explore the possibility that sub GeV dark matter cooled the gas through velocity-dependent, Rutherford-like interactions. 
We argue that such interactions require a light mediator that is highly constrained by 5th force experiments and limits from stellar cooling.  Consequently, only a  hidden or the visible photon can in principle mediate such a force. 
Neutral hydrogen thus plays a sub-leading role and the cooling occurs via the residual free electrons and protons. 
We find that these two scenarios are strongly constrained by the predicted dark matter self-interactions and by limits on millicharged dark matter respectively. 
We conclude that the 21-cm absorption line is unlikely to be the result of gas cooling via the scattering with a dominant component of the dark matter.  An order 1\%  subcomponent of millicharged dark matter remains a viable explanation.
\end{abstract}

\maketitle

\section{Introduction}
A recent measurement of the global 21-cm spectrum by the EDGES collaboration shows a strong absorption signal from  around a redshift of $z=17$~\cite{Bowman2018}. When compared to the standard model prediction, the significance of the excess was estimated to be $3.8 \sigma$.

This first-of-a-kind measurement is an intriguing one.  At around $z\simeq20$ the first stars are born and the cosmic gas in the universe is at its coolest period, before being heated by X-ray radiation. This epoch (roughly $15\lesssim z \lesssim 35$) is known as the cosmic dawn.
The hydrogen 21-cm  transition measurement, as it is usually interpreted, is a unique probe of the temperature of the hydrogen gas.  As the gas decouples from the CMB  at around $z\simeq 200$, its temperature evolves adiabatically, dropping below that of the radiation.   During that period,  modifications of the cosmic history may leave a measurable imprint on the corresponding 21-cm absorption spectrum.

At that same epoch, dark matter (DM) has not yet been stirred up by non-linear gravitational collapse and it is consequently at its coldest phase too. As was first pointed out in~\cite{Tashiro:2014tsa}, DM elastic scatterings with baryons at around or prior to that time, may cool down the gas, thereby influencing the 21-cm absorption spectrum. This idea was further studied in~\cite{Munoz:2015bca}, and recently analyzed in~\cite{Barkana2018} showing that a light (sub-GeV) DM which scatters  off baryons, can leave the desired imprint once Ly-$\alpha$ radiation turns on, providing an exciting explanation of the signal. Such interactions, however, must compete with the Compton scatterings, that acts to couple the gas to the CMB radiation, and hence must be quite strong. A strong velocity-dependence of DM interaction with the SM seems like the best way to evade present day, astrophysical and cosmological constraints~\cite{Jaeckel:2010ni}, including the very recent revisit of the SN1987 limit~\cite{Chang:2018rso}, the BBN and CMB bounds constraining the number of relativistic degrees of freedom~\cite{Davidson:2000hf,Vogel:2013raa}, CMB anisotropy~\cite{Dubovsky:2003yn,Dolgov:2013una,Dvorkin:2013cea,Xu:2018efh}, 5th force experiments~\cite{Adelberger:2003zx,Williams:2004qba,Kapner:2006si,Schlamminger:2007ht,Murata:2014nra}, stellar cooling~\cite{Hardy:2016kme,An:2014twa}, bounds on DM self-interactions~\cite{Tulin:2012wi,Tulin:2013teo}, and bounds on DM millicharge~\cite{Prinz:1998ua,Davidson:2000hf,McDermott:2010pa}.  Accordingly, these interactions must be mediated by a light degree of freedom that allows for Rutherford-like scattering with baryons, $\sigma = \hat{\sigma} v_{\rm rel}^{-4} $, where $v_\mathrm{rel}$ is the relative velocity between the two interacting particles.

The approach taken by~\cite{Tashiro:2014tsa, Munoz:2015bca, Barkana2018} as well as in CMB studies of DM-hydrogen interaction~\cite{Dvorkin:2013cea,Gluscevic:2017ywp, Boddy:2018kfv,Xu:2018efh} is a model-independent one, where one assumes the above velocity dependence in the cross section without specifying its origin.
 In light of recent advancements  it is natural to ask whether there exists a particle physics model that can address the EDGES observation while being consistent with the existing limits.  In this paper we address this question, arguing that the dominant component of DM cannot explain the EDGES observation via the cooling of the hydrogen gas.

To understand this, we note that the de-Broglie wavelength of a DM particle with $m_{\rm DM} \lesssim \units{GeV}$ at $z=20$ is $\lambda\gtrsim10^{-9}\text{ m}$ and always larger than the atomic Bohr radius, $a_0=(\alpha m_e)^{-1}$. Therefore DM interacts with the hydrogen atom as a whole.   In order to have a $v_{\rm rel}^{-4}$-enhanced scattering cross-section, its overall charge must not vanish.  However, that very same property implies that this mediator induces a long range force.  Indeed, the mediator mass, $m_\phi$, must be smaller than the typical momentum transfer in order to induce a $1/v_{\text{rel}}^4$ enhancement. Since the relative velocity at the cosmic dawn is $v_{\text{rel}}\lesssim 10^{-6}$ one finds $m_\phi \lesssim \units{keV}$ for $m_{\rm DM} < \units{GeV}$.  For such a light mediator, as will be demonstrated in Sec.~\ref{sec:models},  constraints from 5th-force experiments are incredibly strong~\cite{Murata:2014nra}  (except for the upper allowed region where  stellar cooling constraints are stronger), ruling out the possibility of cooling of the gas via neutral hydrogen interactions.  Similar constraints hold for any mediator under which heavier atoms are charged (including $U(1)_{B-L}$).  

Strong Rutherford-like interactions between the DM and the gas can still be present due to the small residual fraction of free electrons and protons.  The required interaction rate is roughly three orders of magnitude larger than the one for DM-hydrogen. In the early universe, the interactions with the protons dominate the interaction rate, which however implies an even larger DM-electron cross section.  These cross sections are expected to be probed soon with upcoming direct detection experiments~\cite{Tiffenberg:2017aac, Agnese:2017jvy,SENSEI:2018}.

With the above discussion, two possibilities for the mediation of the DM-gas scatterings remain: A $U(1)_D$ gauge boson (hidden photon) that kinematically mixes with the Standard Model (SM) photon, or the SM photon itself with a DM millicharge.  The  massless limit of the former implies that DM is millicharged under electromagnetism, however strictly speaking the two theories are not the same:  a hidden photon can induce strongly constrained DM self-interactions (see e.g.~\cite{Tulin:2013teo}) while a millicharged DM is strongly constrained as it is expected to be evacuated from the galactic disk~\cite{Chuzhoy:2008zy,McDermott:2010pa,Barkana2018}.  Below we analyze these cases concluding that {\it the cooling of the hydrogen gas via scattering with the dominant component of DM is unlikely to be the explanation of the EDGES observation}.

The paper is organized as follows.   In Section~\ref{sec:21cm} we briefly review the standard model 21-cm physics and discuss the cooling of the gas through its scattering with DM.  In Section~\ref{sec:signal} we present the  DM parameter space required to address the measured absorption line.  Here we account for the fact that DM may interact with the neutral hydrogen, helium, or the free electron and proton fraction.   In Section~\ref{sec:models} we then study the hidden photon and millicharge DM models, demonstrating the strong constraints and identifying the possibly interesting regions for a sub-component of DM.   We conclude in Section~\ref{sec:conclusion}.

\section{The 21 cm global spectrum}
\label{sec:21cm}
\subsection{The Standard Model}
By the time of recombination ($z\sim1100$) matter dominates the energy density in the Universe.
The baryon number density is mainly composed of neutral hydrogen atoms (H), together with a smaller Helium ($\text{He}$) component, $x_{\rm He}=n_{\rm He}/n_{\rm H}\simeq 1/13$, and a small percentage of free protons and electrons, $x_e=n_e/n_{\rm H}=n_{p}/n_{\rm H}$, which varies from $\sim 20\%$ at $z=1100$ to $\sim 2\times 10^{-4}$ at $z\sim20$~\cite{AliHaimoud:2010ab}. This small fraction leaves a crucial imprint in the gas thermal history by coupling $T_{\text{gas}}$ to $T_{\text{CMB}}$ via Thomson scattering down to $z\simeq 200$. 

 Below $z\sim200$ and down to $z\sim30$, the gas is decoupled from radiation and cools adiabatically.   At this time,  most of the hydrogen gas is in its ground state, whose degeneracy is only broken by the hyperfine splitting of the singlet (0)  and triplet (1) states with an energy difference of  $E_{21}=5.9\times 10^{-6}~\rm{eV}\simeq 0.068\text{ K}\simeq 2 \pi/21~\rm cm$~\footnote{Here and below we work in natural units, $\hbar=c=k_B=1$.}.
The relative number density of triplet and singlet states of the hydrogen defines the so called {spin temperature},
\begin{equation}
\frac{n_1}{n_0}\equiv \frac{g_1}{g_0} e^{-E_{21}/T_s}\simeq 3\left(1-\frac{E_{21}}{T_s}\right)\ .
\end{equation}
This effective temperature is sensitive to different spin-flipping processes after recombination and has been extensively studied as an interesting tracker of the cosmological history after recombination~\cite{Field1958,Scott1990,Gnedin:2003wc,Zaldarriaga:2003du,Loeb:2003ya,Barkana:2004zy,Barkana:2004vb,Hirata:2005mz,Pritchard:2006sq,Cooray:2006km,Lewis:2007kz,Pritchard:2008da,Pritchard:2010pa,Morandi:2011hn,Tashiro:2014tsa,Munoz:2015eqa,Munoz:2015bca}.

Roughly speaking, three competing effects influence the evolution of the spin temperature:
\begin{enumerate}
\item
H-H and H-e collisions in the gas induce  $1\leftrightarrow 0$ transitions. The rate can be written as $C_{10}=n_H(k_{10}^H +k_{10}^e x_e + k_{10}^{\rm He} x_{\rm He})$  where $k_{10}^i$ 
are a function of the temperature~\footnote{The rate of H-H collisions dominates over  e-H collisions even though the latter have a much larger cross section. We refer to Ref.~\cite{0004-637X-622-2-1356} for the specific rates we used in our study.}. Since the collision rate $C_{10}$ is much larger than Hubble, $1\to0$ and $0\to1$ collision processes equilibrate
  and one can write 
\begin{equation}
C_{01}=\frac{g_1}{g_0} C_{10} e^{-E_{21}/T_{\text{gas}}}\simeq 3 ~C_{10}\left(1-\frac{E_{21}}{T_{\text{gas}}}\right)\ .
\end{equation} 

\item
Cosmic hydrogen can  resonantly absorb and emit  
the CMB flux, the rates of which are described by the induced emission and absorption rates $B_{10}=B_{01}/3=A_{10}\times T_{\rm CMB}/E_{12}$ which are related to the 
Einstein coefficient describing the spontaneous emission rate, $A_{10}\approx 2.9\times10^{-15}{\rm\ sec^{-1}}$~ \cite{Rybicki1979}.

\item  
Scattering of UV photons can also induce spin-flip transitions of the hydrogen ground state. In particular once absorptions and emissions of Ly-$\alpha$ photons from the first stars 
become important, they
 couple the spin temperature to that of the gas via the Wouthuysen-Field effect~\cite{Wouthuysen1952,Field1958}. We refer the reader to~\cite{Furlanetto:2006jb} for a more detailed explanation of these effects.  
\end{enumerate}

Given that the rates of the processes described above are  larger than Hubble all the way below  $z\lesssim10$, the spin temperature can be defined at any $z$ by an equilibrium equation~\cite{Field1958}\footnote{Often, a different notation is used for this equation.  See e.g.~\cite{Furlanetto:2006jb}.}, 
\begin{equation}
\Delta T_s\simeq \frac{y_{\rm col} \ \Delta T_{\rm gas} + y_{\rm Ly\alpha} \ \Delta T_{\rm Ly\alpha} }{1+ y_{\rm col} + y_{\rm Ly\alpha}}\,,
\label{eq:spinT}
\end{equation}
where $\Delta T = T - T_{\rm CMB}$, and
\begin{equation}
\label{eq:rs}
y_{\rm col} = \frac{E_{\rm 21}}{T_{\rm gas}}\frac{C_{10}}{A_{10}} \quad , \quad y_{\rm Ly\alpha} = \frac{E_{\rm 21}} {T_{\rm Ly\alpha}}\frac{L_{10}}{A_{10}}\,.
\end{equation}
Here $T_{\rm Ly\alpha}$ is defined through the detailed balance equation for the excitation rate via the scattering of Ly-$\alpha$ radiation, $L_{10}$.  The latter is relevant only once stars are formed at around $z\simeq 20$.

Eq.~\eqref{eq:spinT} nicely demonstrates the evolution of the spin temperature relevant for the 21-cm physics. 
At early times, down to $z\sim100$,  collisions dominate, $y_{\rm col}\gg 1,y_{\rm Ly\alpha}$,  and the spin temperature is that of the gas.
 At later times, CMB-induced absorptions followed by emissions begin to dominate, $y_{\rm col, Ly\alpha} \rightarrow 0$, and the spin temperature rises above the gas temperature and towards that of the CMB. Finally, when the Ly-$\alpha$-induced collisions are largest, $y_{\rm Ly\alpha}\gg 1, y_{\rm col}$, at around $z\simeq20$, the spin temperature follows $T_{\rm Ly\alpha}$.
Since the Ly-$\alpha$ radiation can only be hotter than the gas and since the gas is colder than the CMB temperature, one concludes that under the detailed-balance assumptions above, the spin temperature cannot be lower than that of the gas.    This assumption plays an important role when estimating the deviation of the EDGES measurement from standard cosmology.

\subsection{Dark Cooling}\label{sec:dark}

The measured 21-cm  signal to be discussed in the next section is  directly proportional to the difference between the spin and CMB temperatures.  Dark matter can, in principle, cool the spin temperature in two ways: 
\begin{itemize}
\item By scattering off gas particles, DM can allow for an energy flow from the hotter baryonic gas to the cooler dark matter fluid.   Once the spin temperature couples to that of the gas, its temperature is reduced. 
\item DM can directly drain the spin temperature via, for example, direct  spin-flip interactions (for sufficiently light DM) or through bosonically-enhanced induced emissions.   
\end{itemize}
Dark matter cooling of the gas  was first realized in~\cite{Tashiro:2014tsa} and further studied in~\cite{Munoz:2015bca} and~\cite{Barkana2018}.
In this letter we focus on this case, leaving the second possibility for an upcoming publication.
Below we briefly describe the physics involved, referring the reader to~\cite{Munoz:2015bca} for a more detailed analysis.

 Since DM is significantly colder, it is naively expected to cool the gas down through its interaction.  However, the predicted relative bulk velocity between the two gases dissipates with time and thus, under certain conditions, act   to heat up the gas~\cite{Dvorkin:2013cea, Munoz:2015bca}.  
  This competing effect is best seen through the Boltzmann equations describing the evolution of temperatures and of the relative velocity:
\begin{align}
&\frac{dT_{\chi}}{d \log a}=-2 T_{\chi}+\frac{2}{3}\frac{\dot{Q}_{\chi}}{H}\label{eq:Tchi}\,,\\
&\frac{dT_{\text{gas}}}{d \log a}=-2 T_{\text{gas}}+\frac{\Gamma_C}{H}(T_{\text{CMB}}-T_{\text{gas}})+\frac{2}{3}\frac{\dot{Q}_{\text{gas}}}{H}\label{eq:Tgas}\,,\\
&\frac{dv_{\text{rel}}}{d \log a}=-v_{\text{rel}}-\frac{D(v_{\text{rel}})}{H}\,.
\label{eq:vrel}
\end{align}
Here $a$ is the scale factor, $H(z)\simeq\sqrt{\Omega_m} H_0(1+z)^{3/2}$ is the Hubble parameter during matter domination, $v_{\text{rel}}$ is the relative velocity between DM and the gas, $\Gamma_C$ is the Compton scattering rate, and $D(v_{\text{rel}})$ is the drag term that accounts for the relative velocity change due to DM-gas interactions. In the above $\dot Q_{\chi,\rm gas}$ describes the heat transfer per unit time  which results from the scattering of the DM with the gas.  We refer to App.~\ref{sec:heating_formalisem} for a derivation of the drag term and the heat transfer rate.

Originally only interactions with hydrogen were taken into account.  However, as was argued in the Introduction and will be discussed below, free electrons and protons play a crucial role given the severe constraints on the mediators which can induce velocity-enhanced DM-hydrogen interactions.
For the analysis below we therefore consider the  terms describing the interactions with hydrogen, helium and free electrons and protons.  We thus write the different contributions to the heat transfer rate as
\begin{equation}\label{eq:Qdot}
	{\dot Q}_{\rm gas} = \sum_{I=\{\text{H},\rm He,e,p\}}\dot{Q}_{\text{gas}}^I\, .
\end{equation}
Each contribution may be approximated as $\dot{Q}_{\rm gas}^I\sim x_I\Gamma^I\Delta E_I$ with $\Gamma^I \sim n_\chi\sigma^I v_{\rm rel}$ and $\Delta E^I\sim  \mu_{ I} v_{rel}^2$, (where $\mu_{I}$ is the reduced mass of the corresponding DM-gas component, $x^I\equiv n_I/n_{\rm H}$ and $n_\chi$ is the DM number density). Eq.~\eqref{eq:Tgas} then simply explains the need for a large cross-section in order to affect the gas evolution:  In order to cool down the gas efficiently, the rate of DM-gas interactions should be comparable to the heating by Compton scattering $\Gamma_C$. Indeed by requiring  $\dot{Q}_{\text{gas}}\sim\Gamma_C T_{\text{CMB}}$ one can ballpark the required cross section $\hat\sigma^I$ that is needed to explain the 21-cm global spectrum. For example requiring $\dot{Q}_{\text{gas}}\sim\Gamma_C T_{\text{CMB}}$ at $z=20$ with a GeV DM particle, one finds that $\sigma^{H}\simeq10^{-19}\text{ cm}^2$ when assuming that only  DM-hydrogen interactions are switched on.

If DM is a new fundamental particle obeying the basic rules of relativistic quantum field theory, its cross section can grow at small relative velocity at most as $v_{\rm rel}^{-4}$, corresponding to a Coulomb-like  force. In order to enhance as much as possible the cross section at low velocities, below we follow~\cite{Tashiro:2014tsa,Munoz:2015bca} and assume 
\begin{equation} 
\label{eq:xsec}
\sigma^I=\hat{\sigma}^I v_{\rm rel}^{-4}\,.
\end{equation}
This represents the best case scenario for the dark cooling to be enhanced and accommodate a large cross section required to affect the 21-cm spectrum, while not violating CMB and direct detection constraints~\footnote{This argument ignores stronger-than-Coulomb forces that may possibly arise in non relativistic effective theories where DM interactions with matter exhibit a MOND-like behavior \cite{Famaey:2017xou}.}. The root mean square relative velocity between the DM and the gas is ~\cite{Tseliakhovich:2010bj} $v_{\rm rel} = 29\units{km/sec}\sim 10^{-4}$ at decoupling and it redshifts correspondingly at later times.

For the case of light dark matter much below the GeV mass scale (where sufficient cooling can be obtained), the drag force
 has very little effect on the gas temperature evolution and can be neglected.  Assuming the cross section of Eq.~\eqref{eq:xsec}  one then finds~\cite{Tashiro:2014tsa,Munoz:2015bca} the approximated expression,
\begin{equation}
	\dot{Q}_{\text{gas}}^I \simeq \sqrt{\frac{2}{\pi}}\frac{\mu_I}{m_I+m_p}\frac{x^I}{u_{\rm th}^3}\left(T_\chi-T_{\rm gas}\right)n_\chi\hat{\sigma}^I\, ,
\end{equation}
where we defined $(u^I_{\rm th})^2=T_{\rm gas}/m_I+T_\chi/m_\chi$. The corresponding $\dot {Q}_\chi$  is obtained by exchanging $\chi\leftrightarrow\units{gas}$. As can be seen from Fig.~\ref{fig:signal}, the approximate formula above gives a correct description of the behavior of the cooling rate for light dark matter, where the cross section becomes independent on the dark matter mass. 

In the upcoming section we solve these equations more precisely, fitting the observed signal to a DM-gas interaction.

\section{The 21-cm Signal}\label{sec:signal}

Whenever $T_s < T_{\rm CMB}$, the gas absorbs CMB radiation, leaving an imprint in the form of an absorption line in the CMB spectrum. Two lines of that nature are expected to appear, first around $z\sim50-100$ during the dark ages, when collisions dominate the spin-flipping transitions, and second during the cosmic dawn at around $z\sim20$, when the UV radiation from the first stars couples the spin temperature back to the gas. To quantify the absorption strength one defines the redshifted brightness temperature~\cite{Madau:1996cs},
\begin{equation}
\label{eq:Tbright}
T_{21} = \frac{1}{1+z} \left(T_s-T_{\rm CMB} \right)\left(1-e^{-\tau}\right)\,,
\end{equation}
where $\tau$ is the optical depth given by,
\begin{equation}
	\tau\simeq\frac{3 \lambda_{21}^2A_{10}n_{\rm H}}{16T_sH(z)}\,.
\end{equation}    

At $z=17$ the standard evolution described in Sec.~\ref{sec:21cm} predicts $T_{\rm gas}(z=17) \simeq 6.8\units{K}$ (see e.g.~\cite{Tseliakhovich:2010bj}).   Assuming the most optimal scenario, $T_s = T_{\rm gas}$, one then arrives at the brightness temperature,
\begin{equation}
\label{eq:T21SM}
T_{21}^{SM}(z=17) \gtrsim -220 \textrm{ mK}\,.
\end{equation}
This is contrasted with the recent  study by EDGES where
\begin{equation}
\label{eq:T21EDGES}
T_{21}^{EDGES}(z\simeq17) = -500^{+200}_{-500} \textrm{ mK}
\end{equation}
is reported~\cite{Bowman2018},  with  the errors correspond to the $99\%$ C.L.~intervals.    Again, under the assumption of $T_s=T_{\rm gas}$, the above implies $T_{\rm gas}(z=17) = 3.26^{+1.94}_{-1.58}\units{K}$.     The discrepancy between expected and measured temperatures correspond to a 3.8$\sigma$ excess~\cite{Bowman2018}. We are therefore motivated to investigate the possibility that DM-gas interaction underlies the low gas temperature.

\begin{figure}[t]
\centering
\includegraphics[width=0.48\textwidth]{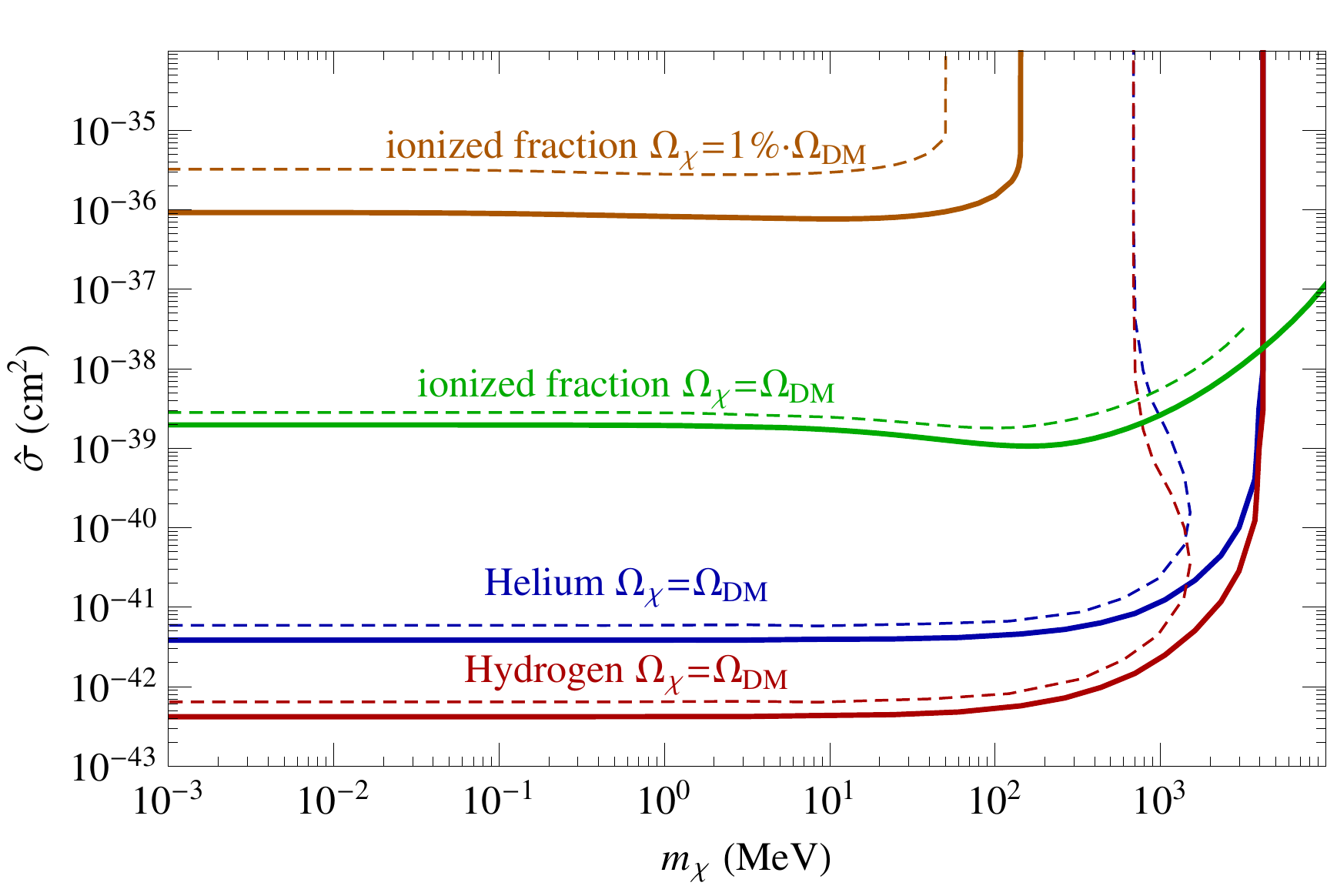}
\caption{The cross section defined in Eq.~\eqref{eq:xsec} required to fit the EDGES signal for DM-hydrogen interactions ({\bf red}), DM-helium interactions ({\bf blue}) and  interactions with the ionized fraction assuming the interacting particle constitutes all of the DM ({\bf green}) or  only 1\% of the DM density ({\bf brown}).  The solid (dashed) lines correspond to the \emph{minimal} cross section needed to obtain a brightness temperature $T_{21}=-300\units{mK}\ (-500\units{mK})$ assuming infinite Ly-$\alpha$ radiation rate which couples the spin temperature to that of the gas, and assuming no heating of the gas due to X-ray radiation.}
\label{fig:signal}
\end{figure}

Solving the full dynamical evolution Eqs.~\eqref{eq:Tchi}--\eqref{eq:vrel} from recombination down to $z=10$ (including the evolution of the free electron fraction), one can extract the necessary DM cross section, $\hat{\sigma}^I$,  needed to produce the reported absorption peak through interaction with a given component of the gas. This analysis was first performed in~\cite{Barkana2018} for the DM-hydrogen interactions, and is now being generalized for DM interaction with the free electron and proton components. The required cross section $\hat{\sigma}$ is shown in Fig.~\ref{fig:signal} as a function of the DM mass. The lines correspond to the minimal cross section that induces an absorption line in agreement with the data at the $99\%$ confidence level.  We choose the parameters in order to give the minimal required cross section between the DM and the gas. In particular any heating effect of the gas at late times from UV radiation or other astrophysical sources is neglected.

Fig.~\ref{fig:signal}   shows that the DM mass needed to explain the signal needs to be lighter than a few GeV. For heavier masses the cross section rises steeply until it becomes impossible for DM interactions to account for the cooling of the gas. This is in agreement with the results of Ref.~\cite{Munoz:2015bca} where it was demonstrated that for DM heavier than the hydrogen, the drag dominates the collisions and causes heating of both DM and the gas.  Since free streaming bounds often require the DM mass to be heavier than a few keV's \cite{Yeche:2017upn,Irsic:2017ixq}, the DM mass region between keV and GeV is the natural window to explain the EDGES excess.  

Our results 
 show that the cooling through interactions with the helium or the ionized fraction requires a much bigger cross section compared to the one for hydrogen.
This can be understood from a simple scaling relation which explains pretty well the behavior of the different cross sections in Fig.~\ref{fig:signal} far away from their turn over at high DM mass. For the case when $m_{\chi}\ll m_p$ and $T_{\chi}\ll T_{\rm gas}$, by inspecting the ratio between the $I$'th component and the hydrogen while holding the cooling rate $\dot Q$ fixed,  Eq.~\eqref{eq:Qdot} implies,
\begin{equation}
	\frac{{\hat \sigma}^I}{{\hat \sigma}^{\rm H}}=\left(\frac{m_I+m_\chi}{m_p+m_\chi}\right)^{2}\left(\frac{m_p}{m_I}\right)^{5/2}\frac{1}{x^I}.
\label{eq:sigratio}	
\end{equation}
 The fact that $x_{H_e}\simeq1/13$ and $x_e\simeq 10^{-4}$ between $z=200$ and $z=20$ helps explain our results.

\section{Models of Dark Matter and Constraints}\label{sec:models}

We now study the circumstances under which viable DM interactions with the SM can explain the EDGES excess.   Several assumptions are made:
\begin{itemize}
\item Dark matter must cool the temperature of the gas.
\item Interactions with the gas occur via a Coulomb-like potential that results in a cross section proportional to  $v_{\rm rel}^{-4}$. 
\item Dark matter is heavier than a keV.
\end{itemize}
Other possibilities that go beyond these assumptions will be presented in an upcoming publication.

The differential cross section for DM scattering with the $I$'th component of the gas can be conveniently parametrized as 
\begin{equation}\label{eq:dsigma_domega}
	4\pi \frac{d\sigma^I}{d \Omega}= \bar\sigma_I  \left|F_{\chi}(q^2)\right|^2 \left|f_I(q^2)\right|^2\,,
\end{equation}
where
\begin{equation}
\label{eq:sigmabar}
\bar\sigma^I = \frac{16\pi \alpha_D\alpha_{\text{eff}}^I~\mu^2_I}{(m_\phi^2+q_{\rm typ}^2)^2}\,,
\end{equation}
not to be confused with $\hat\sigma_I$ defined in Eq.~\eqref{eq:xsec}. Above $m_\phi$ is the mass of the force mediator that couples to the DM with strength $\alpha_D$ and to the SM with strength $\alpha_{\text{eff}}^I$. $F_\chi(q^2)$ is the momentum-dependent DM form factor and $f_I(q^2)$ is the target form factor accounting the the finite size of the target in the case of hydrogen or helium. 
$q_{\rm typ}$ is the typical momentum relevant for the studied scattering process, and is chosen according to the situation.  In recent years it has been established that light, sub-GeV, dark matter may be detected directly via inelastic scattering processes (see, e.g.~\cite{Essig:2011nj,Essig:2012yx,Essig:2015cda,Bloch:2016sjj,Hochberg:2016ajh, Schutz:2016tid,Hochberg:2016ntt,Budnik:2017sbu,Essig:2017kqs,Knapen:2016cue,Knapen:2017ekk,Hochberg:2017wce,Agnes:2018oej}).  For such direct detection experiments the typical momentum transfer is of order the Bohr radius, and therefore in order to make contact with these studies, in what follows we take, $q_{\rm typ} = a_0^{-1}=\alpha m_e$.    

In direct detection experiments, a minimal momentum transfer must exist in order to overcome the experimental threshold.   This is not the case in the early universe and thus for a light mediator, $F_\chi(q^2) \propto 1/q^2$,  the IR divergence of the integral over $d\Omega$ in Eq.~\eqref{eq:dsigma_domega} must be regulated.  As we review in App.~\ref{sec:Cross Sections}, this is done by averaging over the energy transfer, $\dot Q$.  Taking the low mass  limit for the mediator, one finds,
 \begin{equation}
\label{eq:hatdef}
\hat\sigma^I = 
\frac{2\pi\alpha_D\alpha_{\text{eff}}}{\mu_I^2}\log\left(\frac{4\mu^2v_{\rm rel}^2 }{e m_\phi^2}\right)\,.
\end{equation}

As demonstrated in Sec.~\ref{sec:dark}, a velocity dependent cross section is needed in order to enhance it at $z=20$ where the relative velocity between DM and the gas is of order $v_{\text{rel}}=10^{-6}$. By inspection of Eq.~\eqref{eq:dsigma_domega} we note that for that to happen the mediator mass should be smaller than the typical momentum transfer at $z\simeq20$, $q~\sim \mu_I v_{\rm rel}$.  An upper bound on the desired mediator mass is therefore,
\begin{equation}
m_\phi\lesssim \mu_I v_{\text{rel}}\sim \mu_I \cdot 10^{-6}\lesssim 1\text{ keV}\cdot \frac{\mu_I}{1\text{ GeV}}\ ,\label{eq:mediator}
\end{equation}
where in the last inequality we assumed that $m_{\chi}\lesssim 1\text{ GeV}$.   
We now study models that satisfy this criteria.

\subsection{Unscreened Long Range Forces}

\begin{figure}[t!]
\centering
\includegraphics[width=0.48\textwidth]{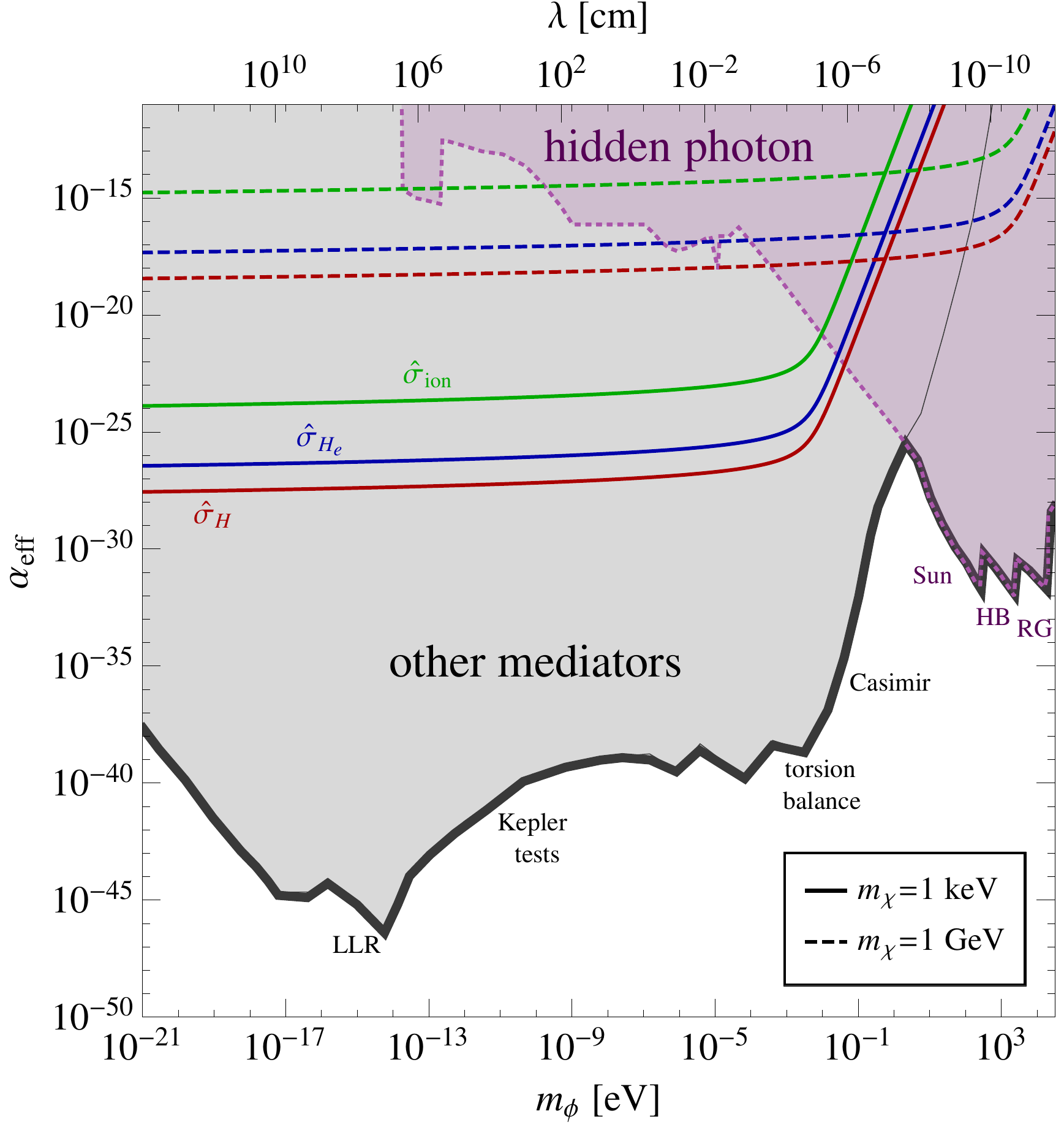}
\caption{Constraints on the effective couplings of light mediators to the SM as a function of the mediator mass.  The {\bf gray shaded} region is excluded for theories that mediate a long range Rutherford-like force which cannot be fully screened. For the astrophisical bounds we assume democratic mediator couplings between electrons and protons. The {\bf purple-shaded} region holds also for a light hidden photon under which SM charges are proportional to their electric charge and can therefore be screened.   The {\bf solid} ({\bf dashed}) lines indicated the \emph{minimal} $\alpha_{\text{eff}}$ needed to fit the EDGES signal (see  Fig~\ref{fig:signal}) when cooling via the scattering of a keV (GeV) DM  with hydrogen ({\bf red}), helium ({\bf blue}) and free electrons and protons ({\bf green}) is assumed.  In order to show the severeness of the constraints on $\alpha_{\rm eff}$, the cross sections are obtained for the best case scenario where the coupling of the DM to the mediator, $\alpha_D=1$, ignoring any possible limits. The gray shaded region is excluded by various 5th force experiments~\cite{Adelberger:2003zx,Salumbides:2013dua}  while the purple-shaded region shows various limits including those from stellar cooling constraints~\cite{Hardy:2016kme,An:2014twa}.
\label{fig:fifthforce}}
\end{figure}

We first consider models where a new light mediator induce Coulomb-like interactions between DM and hydrogen or helium (as apposed to their constituents). In this type of scenarios the mediator mediates a new  unscreened long-range force.  The strength of that force is described by the effective Yukawa potential
\begin{equation} 
V(r)=\frac{\alpha_{\text{eff}}}{r} e^{-m_\phi r}.
\end{equation}
The constraint on the mediator mass given in Eq.~\eqref{eq:mediator} is translated into a minimal  effective range for the force, $(m_\phi)^{-1}\gtrsim 0.1~\mathrm{nm}$.
In Fig.~\ref{fig:fifthforce} we show the limits (total shaded region) on such a mediator in the SM effective  coupling vs. mediator mass plane, alongside the parameters needed to explain the EDGES signal for  keV and GeV DM mass.  The red, blue and green lines indicate the needed couplings assuming gas cooling via hydrogen, helium and ionized fraction respectively.  
We see that, independently of any other constraint related to the particular DM mass or the DM coupling $\alpha_D$, 5th force experiments  alone  severely constrain the coupling of such a light mediatior to the SM~\cite{Adelberger:2003zx,Salumbides:2013dua}. For mediator masses above $0.1\text{ eV}$ the 5th force experiments loose sensitivity and strong limits  from stellar cooling processes in HB stars, in the Sun~\cite{Hardy:2016kme} and in Red Giants~\cite{An:2014twa} take over. In Fig.~\ref{fig:fifthforce} we show the bounds for a mediator which couples to electrons and protons with the same strength. These bounds can be reduced of by a factor of $(m_e/2m_p)^2\approx 10^{-7}$ for mediators coupled to protons only. Even in this advantageous case we see from Fig.~\ref{fig:fifthforce} that the cross sections required to cool the gas in that mass region is excluded by star cooling constraints.

This simple observation rules out models where a new light scalar is the mediator and the effective coupling $\alpha_{\text{eff}}$ is proportional to its Yukawa interactions with a given fraction of the gas $\alpha_{\text{eff}}=y^2_I/4\pi$ (see for example \cite{Knapen:2017xzo}). The same reasoning rules out models with a $B-L$ vector (or scalar) mediator, where  $\alpha_{\text{eff}}=g_{B-L}^2/4\pi$ and the helium carries a charge $2$. The same reasoning apply to models where a light vector mediator is gauging an anomalous symmetry of the Standard Model under which the hydrogen is charged such as $U(1)_B$ or $U(1)_L$.  This last possibility is further constrained by a variety of rare processes and indirect observations related to the existence of anomalous coupling at low energy~\cite{Dobrescu:2014fca,Dror:2017nsg,Dror:2017ehi}.  The line of arguments presented above essentially rules out all known direct DM-atom Coulomb-like scattering as a possible explanation of the EDGES signal.

The 5th force constraints are derived from gravity precision tests. These experiments are  done with electrically neutral systems to reduce the noise. 
Consequently, the only force mediators that may evade the above constraints (and yet  produce a Coulomb-like long-range force) are  those that can be screened at long distances.  Only two such possibilities are known: (i) Models with a 
hidden photon that mixes with the SM photon, and (ii) Models under which DM is millicharged and the mediator is the visible photon itself.
In  both cases, the SM charges under the force are proportional (or equal) to the corresponding electric charge, which is screened at long distances.    We now discuss these two cases in turn.

\subsection{Hidden Photons}
\begin{figure*}[t!]
\centering
\includegraphics[width=0.47\textwidth]{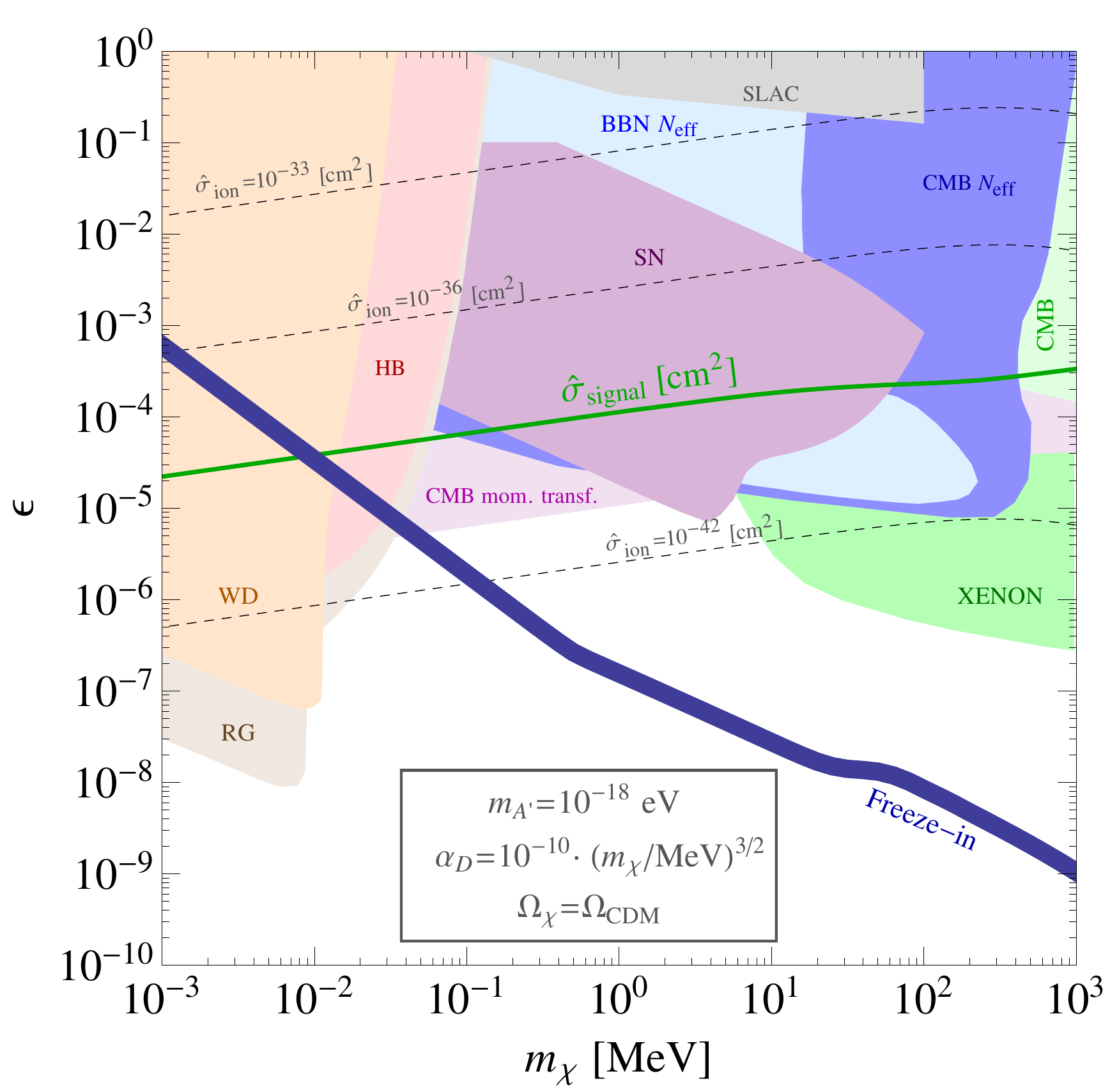}\;\;\;\;\includegraphics[width=0.47\textwidth]{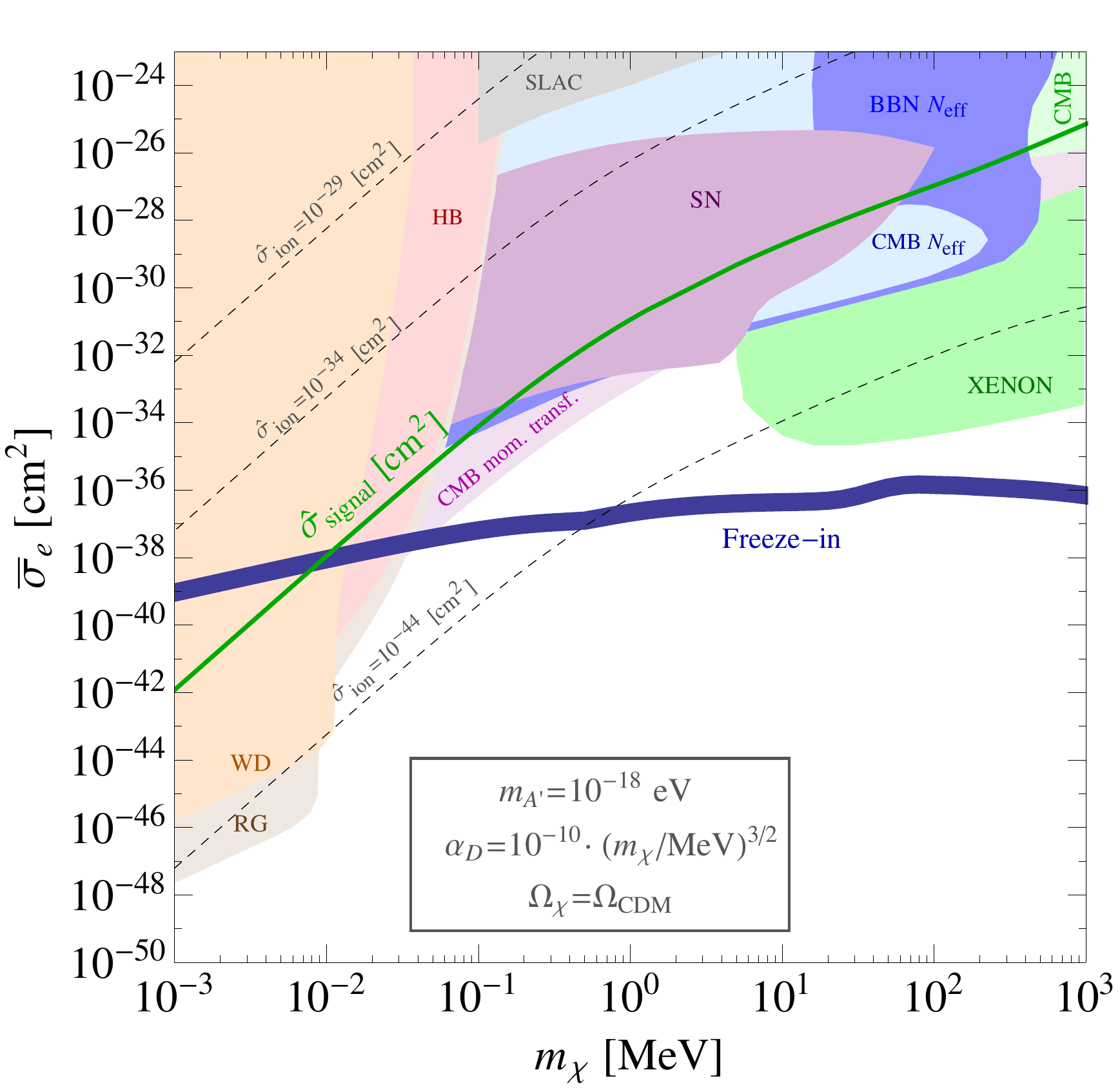}
\caption{Constraints on the hidden photon parameter space for the case when the interacting particle constitutes all of the measured DM density ($\Omega_\chi=\Omega_{\rm CDM}$) and for the choice of very light mediator $m_{A'}=10^{-18} ~\mathrm{eV}$. 
The same parameter space is plotted on the $\epsilon-m_\chi$ plane ({\bf left}) and $\bar{\sigma}_e - m_\chi$ plane ({\bf right}).
 In both plots we choose $\alpha_D=10^{-10}(m_\chi/\mathrm{MeV})^{3/2}$ which is consistent with the self-interaction limits~\cite{Tulin:2013teo}.   The \textbf{green line} represent the minimal cross section needed to explain the EDGES measurement, while in \textbf{dashed-gray lines} we show contours of constant $\hat{\sigma}$ for better orientation.
 Constraints from
cooling of the supernova (SN) 1987A~\cite{Chang:2018rso} ({\bf purple}), direct detection limits from XENON10 \cite{Essig:2012yx,Essig:2017kqs} ({\bf green}),
 effective number of relativistic particles at CMB and at BBN~\cite{Vogel:2013raa} ({\bf blue}), SLAC millicharge experiment~\cite{Prinz:1998ua} ({\bf gray}), cooling of white-dwarfs (WD), horizontal-branch (HB) stars and red-giants (RG)~\cite{Vogel:2013raa} ({\bf pink and brown}), limits on DM-SM coupling at the time of CMB~\cite{McDermott:2010pa} ({\bf light green}) and DM-SM momentum transfer~\cite{Vera_future} ({\bf light purple}) are shown in the shaded regions. 
 \label{fig:darkphoton}}
\end{figure*}

The so-called vector portal  introduces a dark sector that communicates with the SM via a $U(1)_D$ gauge boson, $A^\prime$.  This hidden photon kinetically mixes with the SM hypercharge and consequently with the visible photon,
\begin{equation}
\label{eq:mixing}
{\cal L} \supset -\frac{\epsilon}{2}F^{\mu\nu}F_{\mu\nu}^\prime\,.
\end{equation}
Here $\epsilon$ is the kinetic mixing parameter and $F^{\mu\nu}$  ($F^{\prime\mu\nu}$) is the photon (hidden photon) field strength.  
One may invoke a field redefinition in which the two photons are mass eigenstates that do not mix.  
 For a sufficiently light hidden photon, the form factor of Eq.~\eqref{eq:dsigma_domega} is given by,
\begin{equation}
\label{eq:FF}
F_\chi(q^2) \simeq \frac{\alpha^2 m_e^2}{q^2}\,,
\end{equation}
and the scattering cross sections are given by Eqns.~\eqref{eq:sigmabar} and~\eqref{eq:hatdef} with $\alpha_{\rm eff} = \epsilon^2 \alpha$.

As explained in the Introduction, the de-Broglie wavelength of an interacting sub-GeV DM with a velocity of order 1 km/sec (relevant at $z\simeq 20$) is  too large  to resolve the internal structure of the neutral hydrogen and helium. Consequently,  scattering with these particles does not result in $v_{\rm rel}^{-4}$ interactions. The only possibility left is thus to interact with the free electrons and protons.   The relevant cross section needed in order to address the EDGES measurement is the one shown in green in  Fig.~\ref{fig:signal}. The same conclusion holds for the millicharged DM case discussed below.\footnote{A full treatment of the DM-millicharge interactions has in principle to include the velocity dependent cross section of DM with Hydrogen which we neglect in our approximation $(\lambda\gtrsim a_0)$. Since the Hydrogen is roughly $10^4$ times more abundant at $z\approx20$ dipole interactions of DM with Hydrogen might become relevant for $m_\chi\gtrsim 10\text{ MeV}$. We defer a complete study of this effect for a future work.}

Numerous constraints on the hidden photon and on DM that is coupled to it have been studied in the literature.   One limiting aspect is the self-interaction bounds~\cite{Tulin:2013teo}  which place an upper bound on the DM-mediator coupling, $\alpha_D$.  In Fig.~\ref{fig:darkphoton} we plot the line describing the minimal cross section required to fit the signal in the $\epsilon-m_{\rm DM}$ and $\bar\sigma_e-m_{\rm DM}$ planes.   In the plots we assume the interacting particle constitutes  all of the DM.   On the left, for every value of $\epsilon$ we choose the largest possible $\alpha_D$ coupling allowed by self-interaction limits, taking the mediator mass to be $m_\phi = 10^{-18} \units{eV}$.   
We also show  constraints from
cooling of the supernova (SN) 1987A~\cite{Chang:2018rso}, direct detection limits from XENON10 \cite{Essig:2012yx,Essig:2017kqs},
effective number of relativistic particles at CMB and at BBN~\cite{Vogel:2013raa}, SLAC millicharge experiment~\cite{Prinz:1998ua}, cooling of white-dwarfs (WD), horizontal-branch (HB) stars and red-giants (RG)~\cite{Vogel:2013raa} and limits on DM-SM coupling at the time of CMB~\cite{McDermott:2010pa}. We finally include the most recent limits on DM heating up the gas at the time of CMB (see ~\cite{Vera_future}). These bounds are the dominant ones for DM masses between 100 KeV and 1 MeV. 

It is interesting to ask whether the situation improves if the particle cooling the gas is a subdominant component of DM.  With the exception of the CMB limit (which looses sensitivity for $\Omega_\chi h^2 < 0.007$~\cite{Dubovsky:2003yn}), and the self-interaction bounds, all other constraints remain.  Since $\hat\sigma \propto \epsilon^2\alpha_D$, relaxing the constraint on $\alpha_D$ does not improve or change the best-fit line shown in the $\bar\sigma_e-m_{\rm DM}$ plane.   Thus, excluding a small region around the GeV DM mass and cross section of order 
$\bar\sigma_e \sim 10^{-25}$ cm$^2$, we conclude that the hidden photon mediator cannot explain the observed EDGES signal.

\subsection{Millicharged Dark Matter}

\begin{figure}[t!]
\centering
\includegraphics[width=0.48\textwidth]{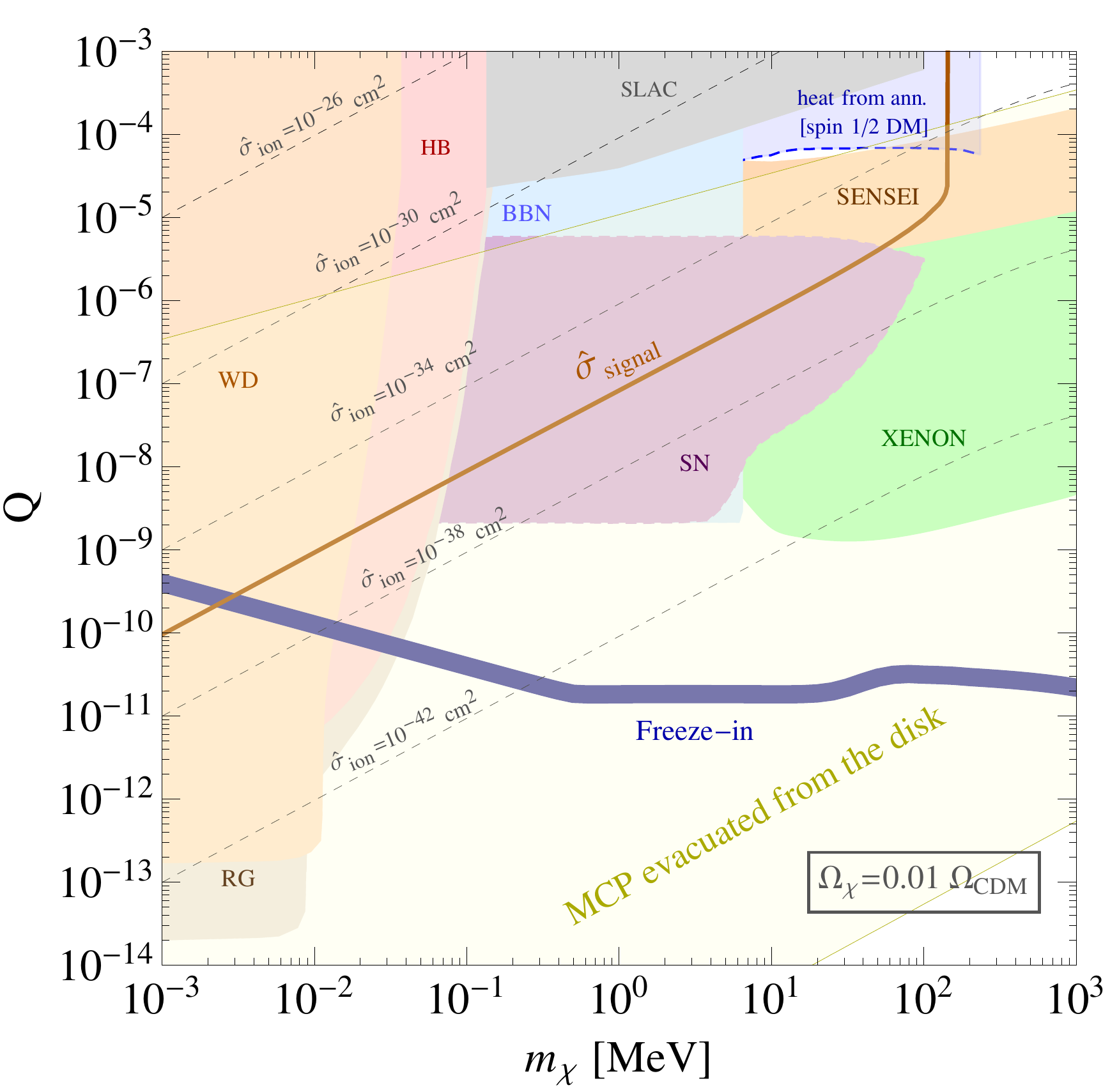}
\caption{Constraints on the charge, Q, of a millicharged particle  as a function of the DM mass.  The {\bf red} line indicates the  minimal cross section needed to explain the EDGES measurement, assuming the millicharged particle constitutes only $1\%$ of the DM density.
 The \textbf{dashed-gray lines}  show contours of constant $\hat{\sigma}$. Constraints from
cooling of the supernova (SN) 1987A~\cite{Chang:2018rso} ({\bf purple}), direct detection limits from XENON10~\cite{Essig:2012yx,Essig:2017kqs} ({\bf green}) and SENSEI~\cite{Crisler:2018gci},
  SLAC millicharge experiment~\cite{Prinz:1998ua} ({\bf gray}), BBN ~\cite{Davidson:2000hf} (\textbf{light blue}) and cooling of white-dwarfs (WD), horizontal-branch (HB) stars and red-giants (RG)~\cite{Vogel:2013raa} ({\bf pink and brown}) are shown in the shaded regions. We also add constraints from heating due to DM annihilation derived in \cite{Liu:2018uzy} ({\bf blue}). This bound only applies to fermionic DM for which the annihilation is $s$-wave. The shaded yellow band  indicates where millicharge DM might be evacuated from the galactic disk \cite{Chuzhoy:2008zy,McDermott:2010pa}. 
}
\label{fig:millicharge}
\end{figure}

Millicharged particle (MCP) DM is comprised of particles that are charged directly under electromagnetism. Severe constraints on the strength of DM-SM interaction   force its electric charge  to be fairly small. The MCP form factor can also be approximated by Eq.~\eqref{eq:FF} and the cross section is given by Eqs~\eqref{eq:sigmabar} and~\eqref{eq:hatdef} with $\alpha_{\rm eff}=\alpha$ and $\alpha_D = Q^2 \alpha$, where $Q$ is the MCP electric charge in units of the electron's electric charge. While in the $m_{A'}\to0$ limit of a kinetically mixed $U(1)_D$ discussed above DM matter appears to carry an electric millicharge, there are still some subtle differences between the two scenarios. In the case of a vector portal, DM-DM interaction are mediated by a hidden photon and  are proportional to $\alpha_D$. Millicharged DM, on the other hand, self-interacts only through the exchange of the SM photon, thus evading most DM self interactions constraints. Moreover, the number of effective degrees of freedom in a  ``pure" millicharge DM is smaller than that of a ultr-light hidden photon. The bounds from BBN and CMB are then relaxed. Millicharged particles might also be evacuated from the galactic disk~\cite{Chuzhoy:2008zy,McDermott:2010pa,Barkana2018}. This effect, if true, exclude MCP from being responsible for the whole DM budget. A precise assessment of this effect including the uncertainties on the modeling of the various components of the Galactic magnetic field is left for future investigation.

In Fig.~\ref{fig:millicharge} we plot the measured EDGES signal on top of the existing constraints in the $Q-m_\chi$ plane, assuming only 1\% of DM is in the form of MCP.  As evident from comparison to Fig.~\ref{fig:darkphoton}, the parameter space slightly opens, the main reason being that BBN and CMB constraints are weaken in the absence of a hidden photon. Weaker cosmological constraints still applies, in particular we include the BBN bound from Ref.~\cite{Davidson:2000hf}. We also show the region where DM annihilation heats up the gas, consequently the spin temperature would rise and thus the EDGES signal can not be accounted for in that region. This effect was first pointed out in \cite{DAmico:2018sxd} and was carefully computed in the case of MCP in \cite{Liu:2018uzy}. Note that this bound is only relevant for fermionic MCP where the thermal-averaged annihilation cross section is velocity independent. One might wonder why the bound computed in \cite{Liu:2018uzy} seems independent on the millicharge $q$ in the region of interest. This exactly the regime in which the heating from annihilation equates the cooling from scattering described in Eq.~(\ref{eq:Tchi}-\ref{eq:Tgas}). Since both these cross section depends on $q^2$ the bound is approximately independent on $q$. 

The constraints from XENON10 and SENSEI were re-scaled to account for the smaller DM fraction and a preliminary first estimate of the terrestrial effect on the charged particle flux penetrating the earth was included~\cite{FutureEssig}.  Indeed this effect to date has only been studied for a much heavier mediator~\cite{Emken:2017erx}. The shaded light region taken from Refs.~\cite{Chuzhoy:2008zy,McDermott:2010pa} is there as a reminder that a significant reduction of the DM flux might be caused by MCP being evacuated from the galactic disk. This region should not be treated as an exclusion region since in Fig.~\ref{fig:millicharge} MCP is only a subdominant constituent of the DM density.

\section{Conclusions}\label{sec:conclusion}
In this paper we studied the possibility that the strong 21-cm absorption line observed by the EDGES collaboration can be explained due to the cooling of the hydrogen gas via its scattering with cold dark matter.   In order to explain the observed signal, dark matter must strongly interact with the gas at around $z=20$, implying that Rutherford-like (velocity-enhanced) interactions must induce the cooling.  

Such scatterings require a very light mediator and two possibilities exist: Either the hydrogen or helium are charged under the new long-range force (meaning that the nucleons and electrons do not screen the interaction) or they are neutral.   In the former case, 5th-force experiments strongly constrain the possibility of mediating the required strong interaction between the DM and the visible sector.   The latter case can arise from either the interaction with the visible photon or with a hidden photon that kinematically mixes with the visible one.    We showed that both of these possibilities are strongly constrained due to limits on millicharge dark matter and self-interacting dark matter respectively.    Consequently, the dominant DM component cannot cool the hydrogen enough to explain the observed signal.
In the case of a millicharged particle, a subcomponent of the DM ($\lesssim 1\%$)  may explain the signal while marginally evading the bounds.

\medskip

\subsection{Acknowledgements}
We  thank Nima Arkani-Hamed,  Yuval Grossman and especially   Matias Zaldarriaga  for many enlightening discussions. We also thank , Rouven Essig and  Jae Hyeok for sharing with us their SN constraints result and preliminary results on terrestrial effects on MCP and their effect on direct detection reach. Finally we thank Hongwan Liu, Tracy Slatyer and Paolo Panci for useful discussions regarding DM annihilation. NJO is grateful to the Azrieli Foundation for the award of an Azrieli Fellowship.  NJO, DR and TV are supported in part by the I-CORE Program of the Planning Budgeting Committee and the Israel Science Foundation (grant No. 1937/12).  DR and TV are further supported by the Israel Science Foundation-NSFC (grant No. 2522/17) and by the German-Israeli Foundation (grant No. I-1283- 303.7/2014).  TV is further supported by the Binational Science Foundation (grant No. 2016153), by the European Research Council (ERC) under the EU Horizon 2020 Programme (ERC-CoG-2015 - Proposal n. 682676 LDMThExp), and by a grant from the Ambrose Monell Foundation, given by the Institute for Advanced Study. For RB, this publication was made possible through the support of a grant from the John Templeton Foundation; the opinions expressed in this publication are those of the author and do not necessarily reflect the views of the John Templeton Foundation. RB was also supported by the ISF-NSFC joint research program (grant No. 2580/17).

\medskip

{\bf Note added}: While this paper was in preparation, the related papers~\cite{Munoz:2018pzp, Berlin:2018sjs} appeared.

\appendix
\section{Heating Formalisem} 
\label{sec:heating_formalisem}
The temperature and relative bulk velocity evolution is described by Eqns.~\eqref{eq:Tgas},~\eqref{eq:Tchi} and~\eqref{eq:vrel} and can be solved once the drag term $D(V_{\rm rel})$ and the heating rate $\dot Q$ are known. The drag term is a consequence of the momentum transfer between the DM and the baryonic gas and is given by,
\begin{equation}
	V_{\rm rel}\,D(V_{\rm rel})=\vec{V}_{\rm rel}\cdot\left(\vec{D}_\chi-\vec{D}_{\rm gas} \right),
\end{equation}
with
\begin{align}
	&\vec{D}_\chi=\frac{n_{\rm gas}}{m_\chi}\left<v_{\rm rel}\int d \Omega \frac{d \sigma}{d \Omega} \vec{q}\right>,\\
	&\vec{D}_{\rm gas}=-\frac{n_\chi}{m_{\rm gas}}\left<v_{\rm rel}\int d \Omega \frac{d \sigma}{d \Omega} \vec{q}\right>.\nonumber
\end{align}
above $\langle\cdot\rangle$ denotes thermal averaging, $\vec{q}$ is the momentum transfer in a single collision, and $v_{\rm rel}$ is the relative velocity between the particles participating in the interaction. The drag term is thus given by
\begin{equation}
	V_{\rm rel}\,D(V_{\rm rel})=\frac{\rho_\chi+\rho_{\rm gas}}{m_\chi+m_{\rm gas}}\left<\frac{v_{\rm rel}}{\mu}\int d \Omega \frac{d \sigma}{d \Omega} \vec{q}\right>\cdot \vec{V}_{\rm rel},
\end{equation}
where $\mu$ is the interacting particles reduced mass. The drag term is Galilean invariant and thus can be computed in the desired frame. 

To calculate the heat transfer we move to the momentarily rest frame of one of the components and calculate the thermal averaged rate of energy transfer, the gas to DM heat transfer is thus given (in any frame) by,
\begin{equation}
	\dot{Q}_{\chi}=n_{\rm gas}\left< v_{\rm rel}\left(\vec{v}_{\rm cm}-\vec{V}_\chi\right)\cdot \int d \Omega \frac{d \sigma}{d \Omega} \vec{q}\right>,
\end{equation}
where $\vec{V}_\chi$ is the dm bulk velocity and $\vec{v}_{\rm cm}$ is the interacting particles center of mass velocity. The DM to gas heat transfer can be obtained by replacing $({\rm gas} \leftrightarrow \chi )$ and $(\vec{q}\to-\vec{q})$. Using the two equations above one can derive the following conservation law
\begin{equation}
		n_\chi\dot{Q}_\chi+n_{\rm gas}\dot{Q}_{\rm gas}=\frac{\rho_\chi \rho_{\rm gas} }{\rho_\chi+\rho_{\rm gas}} V_{\rm rel} D(V_{\rm rel}).
\end{equation}

\section{Cross Sections} 
\label{sec:Cross Sections}
From the above section we learn that both the heat transfer and the drag term are related to the underlying particle physics through the quantity,
\begin{equation}
	\vec{I}=\int d \Omega\frac{d \sigma}{d \Omega}\vec{q}.
\end{equation}
The momentum transfer is a function of the scattering angle and accordingly a function of the solid angle $\Omega$. The only direction in the above quantity is the initial relative momentum $\vec{p}=\mu\vec{v}_{\rm rel}$, this in turn motivates us to define the transfer cross section as,
\begin{equation}
	\sigma_T=\frac{\vec{p}\cdot\vec{I}}{p^2}=\frac{\pi}{2 p^4}\int_0^{4p^2}dq^2\frac{d \sigma}{d \Omega}q^2,
\end{equation}
where for the last equality we made use of the facts that in elastic scattering $\vec{p}\cdot\vec{q}=-q^2/2$ and $d \cos \theta=-dq^2/2p^2$. We note that this definition of the transfer cross section is identical to the one given in \cite{Tulin:2013teo,Gerjuoy1965}.

In cases where the cross section gets a $v_{\rm rel}^{-4}$ enhancement the differential cross section is in the form of Eq.~\eqref{eq:dsigma_domega} with $F_\chi\simeq 1/q^2$ and $f_I\simeq 1$,
\begin{equation}
\label{eq:xsecapp}
	4\pi \frac{d\sigma}{d \Omega}= \frac{16\pi\alpha_D\alpha_{\text{eff}}~\mu^2}{(m_\phi^2+q^2)^2}.
\end{equation}
The total cross section obtained by integration of the above equation is seemingly divergent when the mass of the vectors is taken to zero. However, the transfer cross section of interest to us is physically regularized, to leading order in the mediator mass it is given by,
\begin{equation}
	\sigma_T \simeq \frac{2\pi\alpha_D\alpha_{\text{eff}}}{\mu^2v_{\rm rel}^4}\log\left(\frac{4\mu^2v_{\rm rel}^2 }{e\,m_\phi^2}\right),
\end{equation}

which indeed depend only logarithmically on the mediator mass.   
This cross sections is however only the Born approximation to the total cross section, and holds as long as $m_\phi\gg\epsilon\sqrt{\alpha_D \alpha_{\rm eff}}m_\chi$.
We note that the limit $m_\phi\to0$ is regularized via thermal masses that exists inside the plasma.

			\medskip
\bibliography{21.bib}

\end{document}